\documentclass[aip,pop,groupedaddress,reprint]{revtex4-1}
\usepackage{graphicx}
\usepackage{ulem}
\usepackage{amssymb}
\usepackage{amsmath}
\usepackage{todonotes}
\begin{document}
%\draft                    

%\flushbottom
%\twocolumn[
%\hsize\textwidth\columnwidth\hsize\csname @twocolumnfalse\endcsname

\title{Field-emission from parabolic tips:  current distributions, the net current and effective emission area}
 
%\vskip 0.3 in

\author{Debabrata Biswas}

\affiliation{
Bhabha Atomic Research Centre,
Mumbai 400 085, INDIA \\
Homi Bhabha National Institute, Mumbai 400 094}

%\pacs{85.45.-w}{}
%\pacs{03.65.Sq}{}
%\pacs{03.65.Xp}{}
%\pacs{52.59.Sa}{}

\begin{abstract}
  Field emission from nano-structured emitters primarily takes place from the tips. Using
  recent results on the variation of enhancement factor
  around the apex (Biswas et al, Ultramicroscopy 185, 1-4 (2018)), analytical expressions for surface distribution
  of net emitted electrons as well as the total and normal energy distributions are derived in terms
  of the apex radius $R_a$ and the local electric field at the apex $E_a$.
  Formulae for the net emitted current and  effective emission area in terms of these quantities
  are also obtained. 
\end{abstract}

%\email{dbiswas@barc.gov.in}
%\maketitle

%\pacs{85.45.-w}{}
%\pacs{03.65.Sq}{}
%\pacs{03.65.Xp}{}
%\pacs{52.59.Sa}{}

%\date{\today}
%\vskip 0.2 in
%\centerline{\bf Abstract}

%\vskip 0.25 in

%\pacs{85.35.-p, 03.65.Sq, 52.59.Sa}

\maketitle

%]
\newcommand{\be}{\begin{equation}}
\newcommand{\ee}{\end{equation}}
\newcommand{\bea}{\begin{eqnarray}}
\newcommand{\eea}{\end{eqnarray}}
\newcommand{\Tbar}{{\bar{T}}}
\newcommand{\En}{{\cal E}}
\newcommand{\K}{{\cal K}}
\newcommand{\GC}{{\cal \tt G}}
\newcommand{\Lop}{{\cal L}}
\newcommand{\DB}[1]{\marginpar{\footnotesize DB: #1}}
\newcommand{\q}{\vec{q}}
\newcommand{\kt}{\tilde{k}}
\newcommand{\Lopn}{\tilde{\Lop}}
\newcommand{\noi}{\noindent}
\newcommand{\ovn}{\bar{n}}
\newcommand{\ovx}{\bar{x}}
\newcommand{\ovE}{\bar{E}}
\newcommand{\ovV}{\bar{V}}
\newcommand{\ovU}{\bar{U}}
\newcommand{\ovJ}{\bar{J}}
\newcommand{\calE}{{\cal E}}
\newcommand{\ovphi}{\bar{\phi}}
\newcommand{\zt}{\tilde{z}}
\newcommand{\rt}{\tilde{\rho}}
\newcommand{\tth}{\tilde{\theta}}
\newcommand{\nuv}{{\rm v}}
\newcommand{\ck}{{\cal K}}
\newcommand{\cc}{{\cal C}}
\newcommand{\ca}{{\cal A}}
\newcommand{\cb}{{\cal B}}
\newcommand{\cg}{{\cal G}}
\newcommand{\ce}{{\cal E}}
\newcommand{\fn}{{\small {\rm  FN}}}

%\newpage
%\noindent

%\section{Introduction}
%\label{sec:Introduction}

\section{Introduction}
\label{sec:intro}

Field emission is increasingly a preferred source of electrons in several applications including
vacuum microwave and terahertz devices, microscopy, lithography and space and medical applications.
Theoretical prediction of the field-emission current largely relies
on the Fowler-Nordheim (FN) \cite{FN} model which assumes a planar emission surface in addition to  
the free-electron model.
The Murphy-Good adaptation of the Fowler-Nordheim law serves
as the basis of modern analytical field emission formulae \cite{murphy,Forbes,forbes_deane,jensen2003}.
It includes the image-charge correction and makes allowance for curved emitters
by replacing the  electric field on the planar emitter surface by the (enhanced) local electric field $E({\bf{r}})$ at a point
${\bf{r}}$ on the curved emitter surface.
This quasi-planar formalism serves as a reasonable approximation to the current density
for even  high aspect-ratio curved field emitter  when the local field at the emitter surface
is not too low. The bottleneck in most theoretical predictions
lies in not knowing much about the enhanced electric field $E({\bf{r}})$ or its
local variation. In the absence of such information,
it is impossible to predict the net emission current  or the distribution of electrons
(that are emitted from a single emitter comprising of
the apex and its neighbourhood) with respect to launch angle, the total and normal energy \cite{uncertainties}.

Our aim here is to provide a partial rectification of this lacuna for the class of smooth axially symmetric
curved emitters placed on a planar surface in situations where the asymptotic (away from the
emitter apex) electric field $E_0$  is uniform and aligned parallel to the emitter axis.
Under such circumstances, a recent result \cite{db_ultram}
shows that the local field

\be
E({\bf{r}}) = E_a \cos\tth \label{eq:var}
\ee

\noi
where $E_a = \gamma_a E_0$ is the field at the emitter apex,

\be
\cos\tilde{\theta} = \frac{z/h}{\sqrt{(z/h)^2  + (\rho/R_a)^2}} \label{eq:costhet}
\ee

\noi
and $\gamma_a$ is the field enhancement factor at the apex.
In Eq.~\ref{eq:costhet}, $R_a$ is the apex radius of curvature and $h$ is the height of the emitter-apex measured
from the planar surface. It is assumed that the  curved protrusion acting as a field emitter is aligned in the
direction of the externally applied field (the z-axis) and has a smooth tip described by $z = g(\rho)$
where $\rho^2 = x^2 + y^2$ for a point on the emitter surface.
Thus a Taylor expansion at the apex yields

\be
z = h + \rho \big(\frac{dg}{d\rho}\big)_{|_{\rho = 0}} + \frac{\rho^2}{2}\big(\frac{d^2g}{d\rho^2}\big)_{|_{\rho = 0}} + \ldots \simeq  h - \frac{\rho^2}{2R_a}
\ee

\noi
since $(dg/d\rho)_{|_{\rho = 0}} = 0$. Thus, the tip is locally parabolic. Using Eq.~\ref{eq:var} and assuming
that $E_a$ is known, an ideal theoretical
prediction for field-emission current from a single vertically aligned emitter can in principle be made.

Equation~\ref{eq:var} also allows us to address the issue of distribution of electrons
from a single emitter. A joint distribution of the net emitted electrons with respect
to generalized angle $\tth$, the total energy $\ce_T$ and the normal energy $\ce_N$ can thus be written down immediately
using Eq.~\ref{eq:var} and \ref{eq:FN} while distributions with respect to individual
quantities can be arrived at by integrating over the other two quantities.
For instance, we denote by the quantity $f_S(\tth) d\tth$, the current emitted from
the surface at generalized angle $\tth$
between $\tth$ and $\tth + d\tth$. We refer to this as the surface-angular current
distribution and $f_S(\tth)$ as the corresponding density.
While the tip or apex ($\rho = 0$) has
the maximum field enhancement, the emitted
current from its immediate neighbourhood is clearly negligible as the emission area is small. On the
other hand, as $\tilde{\theta}$ increases, the enhancement factor $\gamma$ decreases while the size
of the area element increases. Thus the surface-angular distribution of emitted electrons must have a peak
close to the apex. We show here that this distribution is universal in the sense that
all emitters with a given apex curvature and local apex field have identical distribution
with respect to $\tilde{\theta}$. We also establish that $\tth$ is in fact the launch angle $\theta_L$
for sharp emitters and thus has a physical significance.

Similarly, the total-energy distribution $f_T(\ce_T)d\ce_T$ and normal-energy distribution
$f_N(\ce_N)d\ce_N$ of emitted electron current
are also of interest \cite{young,gadzuk,liang,egorov}. These are distinct from earlier theoretical studies that deal with 
the energy distribution of  current {\it density} at a {\it single point on the emitter surface}.
The two sets of distributions are thus very different but can be related by a surface integration over the emitter surface
using the variation in field enhancement
factor in the neighbourhood of the apex. We thus study distributions of the total emitted 
current from an emitter rather than the emitted current density at a point on the emitter.

In the following, we shall establish analytical formulae for total emitted current from a parabolic
tip as well as the various distributions mentioned above in terms of $E_a,R_a$ and
$h$. While $R_a$ and $h$ are experimentally measurable quantities, we shall assume
the local field at the apex  $E_a$, to be known as well \cite{emperical}.

\section{The surface (angular) distribution of emitted electrons}

We shall first address the question of surface distribution of emitted electrons. It is assumed that the tip is
smooth having a radius of curvature $R_a > 20$~nm \cite{sharp_tip,BR2017a,BRS2017b}.

In the regime, $R_a > 20$~nm, and for local field strengths less than $10$ V/nm,
the zero-temperature Murphy-Good formulation \cite{murphy} of
Fowler-Nordheim (FN) type field emission formula can be used for the local electron current density,
$J({\bf{r}})$, at a point ${\bf{r}}$ on the emitter surface. It can be expressed as \cite{forbes_deane}

\be
J({\bf{r}}) =  \frac{1}{t_F^2({\bf{r}})} \frac{A_\fn}{\phi} (E({\bf{r}}))^2 
e^{-B_\fn \nuv_F({\bf{r}}) \phi^{3/2}/E({\bf{r}})}. \label{eq:FN}
\ee

\noi
In the above, the free electron model is assumed and barrier lowering due to the image potential is incorporated.
Here, $A_\fn~\simeq~1.541434~{\rm \mu A~eV~V}^{-2}$ and 
$B_\fn~\simeq~6.830890~{\rm eV}^{-3/2}~{\rm V~nm}^{-1}$ are the conventional FN constants,  $\phi$
is the work function while $\nuv_F \simeq 1 - f + \frac{1}{6}f\ln f$ and $t_F \simeq$~$1 + f/9 - \frac{1}{18}f\ln f$ are correction factors due to the
image potential with $f = f({\bf{r}}) \simeq c_S^2 E({\bf{r}})/\phi^2$. The local field
$E({\bf{r}})$ is the magnitude of the local electric field  at the emitter surface while $c_S$ is the
Schottky constant with $c_S^2 = 1.439965~{\rm eV^2~V^{-1}~nm}$.
Note that Eq.~\ref{eq:FN} does not depend on the Fermi energy as it relies on a WKB formula
for the transmission coefficient. This induces errors that have been extensively
tabulated by Mayer \cite{Mayer} for different work function, applied field and Fermi energy.
We shall ignore these subtleties here and merely note that a correction factor to Eq.~\ref{eq:FN}
can be used when absolute or unnormalized quantities are evaluated.

The current emitted from the strip on the emitter surface between radius $\rho$ and $\rho + d\rho$ can
be expressed as

\be
f({\bf{r}}) dr = J({\bf{r}}) 2\pi \rho \sqrt{1 + (dz/d\rho)^2} d\rho
\ee

\noi
where ${\bf{r}} = (\rho,z,\phi)$. For axially symmetric vertically aligned emitters, $z = h - \rho^2/(2R_a)$
near the tip. Thus,

\be
f({\bf{r}}) dr = J({\bf{r}}) 2\pi \rho \sqrt{1 + (\rho/R_a)^2} d\rho.
\ee

\noi
The transformation from $d\rho$ to $d\theta$ where $\tan\theta = \rho/z$ should in principle give
the angular distribution of emitted electrons. However, it is clear that such a distribution
cannot be universal and will depend on the emitter height and radius of curvature. A transformation
to normalized co-ordinates $\tilde{\rho} = \rho/R_a$ and $\tilde{z} = z/h$ is however helpful.
With $\tan\tth = \tilde{\rho}/\tilde{z}$, the surface angular distribution is expressed as

\be
f_S(\tth) d\tth =  J({\bf{r}})~ 2\pi \rho ~\sqrt{1 + \rt^2}~ \frac{d\rho}{d\tth} d\tth
\ee

\noi
where, using $\sin\tth = \rt/\sqrt{\zt^2 + \rt^2}$, we have

\bea
\cos\tth~ d\tth & = & \frac{d\rho}{R_a}\frac{1}{ \sqrt{\zt^2 + \rt^2}} \Big[ 1 - \sin^2\tth \big\{ 1 - \frac{zR_a}{h^2} \big\} \Big] \\ 
& \simeq & \frac{d\rho}{R_a}\frac{1}{ \sqrt{\zt^2 + \rt^2}} \cos^2\tth
\eea

\noi
where we have neglected $zR_a/h^2$ for sharp emitters ($h >> R_a$). Thus

\bea
\frac{d\rho}{d\tth} & \simeq &  R_a \frac{\sqrt{\zt^2 + \rt^2}}{\cos\tth} \\
& \simeq & R_a \frac{\sqrt{1 + \rt^2}} {\cos\tth}
\eea

\noi
since for sharp emitters, emission is limited to regions for which $z \simeq h$ (or $\zt \simeq 1$).
Finally, 

\bea
f_S(\tth) d\tth & = & 2\pi R_a^2 J({\bf{r}})~ \frac{\rt}{\sqrt{1 + \rt^2}} ~\frac{(1 + \rt^2)^{3/2}}{\cos\tth} ~ d\tth \\
& \simeq & 2\pi R_a^2 J({\bf{r}})~\frac{\sin\tth}{\cos^4\tth} d\tth
\eea

\noi
where

\be
J({\bf{r}}) =  \frac{1}{t_F(\tth)^2} \frac{A_\fn}{\phi} E_a^2 \cos^2\tth~
e^{-B_\fn \nuv_F(\tth) \phi^{3/2}/(E_a \cos\tth)}. \label{eq:FN1}
\ee

\noi
Note that $t_F \simeq  1 + f/9 - \frac{1}{18}f\ln f$ and $\nuv_F \simeq 1 - f + \frac{1}{6}f\ln f$
also depend on $\tth$ through $f = c_S^2 E_a \cos\tth/\phi^2$. Putting everything
together, we have

\be
f_S(\tth) d\tth = 2\pi R_a^2~\frac{\sin\tth}{\cos^2\tth} \frac{A_\fn}{\phi}\frac{E_a^2}{t_F^2(\tth)} e^{-\frac{B_\fn \nuv_F(\tth) \phi^{3/2}}{(E_a \cos\tth)}} d\tth. \label{eq:angular}
\ee

\noi
Eq.~\ref{eq:angular} forms the central result of this section. It shows that in normalized
co-ordinates, the angular distribution of emitted electrons is universal for sharp
emitter tips for a given apex local field ($E_a$) and radius of curvature ($R_a$).

\begin{figure}[htb]
\vskip -2.1 cm
%\hskip -1.0cm
%\centering
\hspace*{-1.0cm}\includegraphics[width=0.6\textwidth]{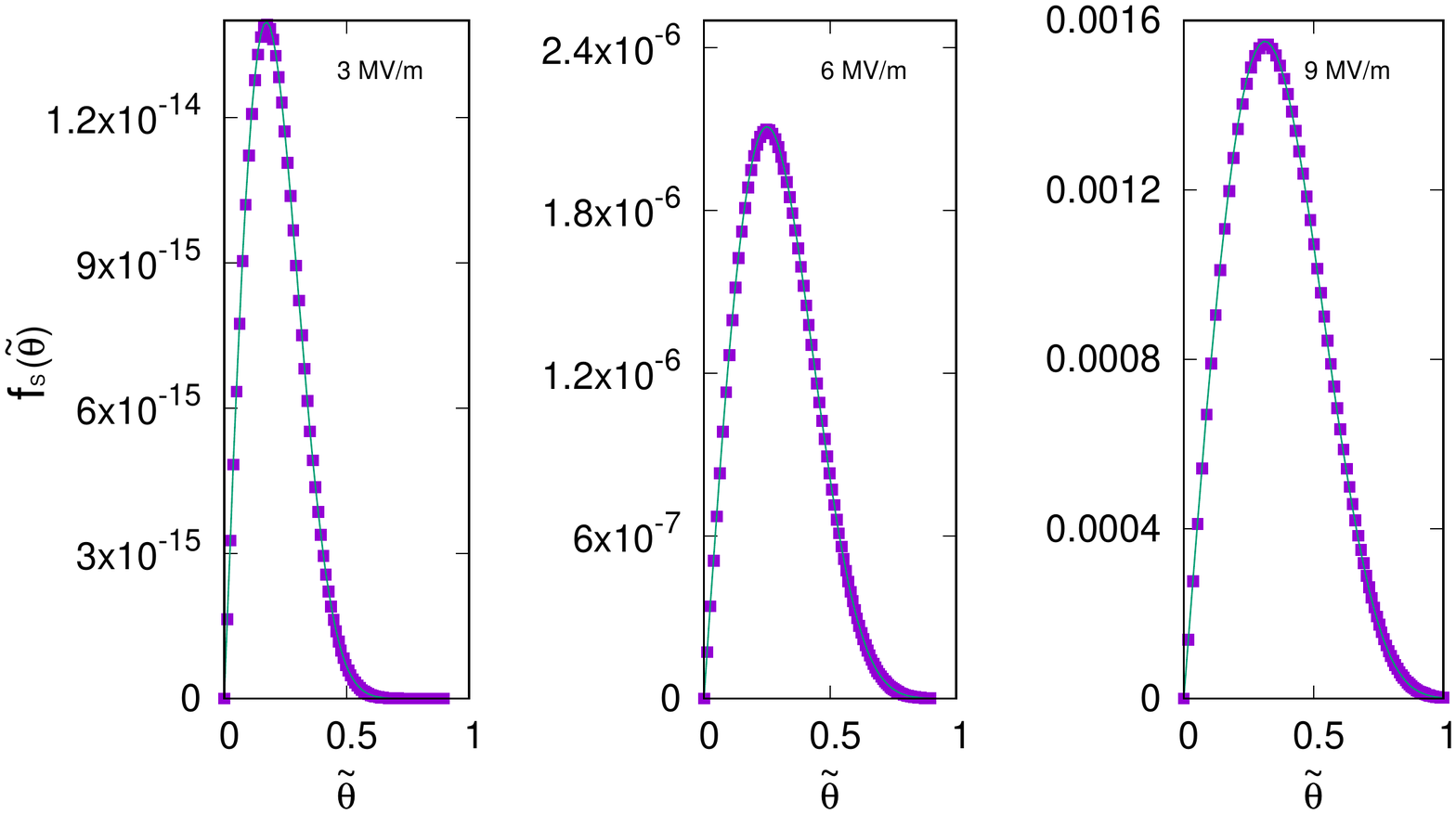}
\vskip -0.6 cm
\caption{The surface angular density of current for an emitter with $\gamma_a = 617$
  for three different asymptotic external fields: (left) $E_0 = 3$~MV/m  (centre) $E_0 = 6$~MV/m and
  (right) $E_0 = 9$~MV/m (corresponding to local apex field values, $E_a$,
  of 1.85~V/nm, 3.70~V/nm, and 5.55~V/nm, respectively).
  The height $h = 1500~\mu$m, the apex radius of curvature
  is $R_a = 0.26~\mu$m and the work function $\phi = 4.5$eV.
  The continuous line is the prediction of Eq.~\ref{eq:angular}. The angle $\tth$ is measured in
  radian.
}
\label{fig:angular}
\end{figure}

Fig.~\ref{fig:angular} shows a typical (un-normalized) angular density of currents for an emitter
with apex radius $R_a = 0.26~\mu$m and height $1500~\mu$m, having an enhancement factor
$\gamma_a = 617$. The work function $\phi = 4.5$~eV and $E_F = 8.5$~eV.
The solid squares are ``exact'' results and have been obtained
using a numerical scheme for the transmission coefficient \cite{DBVishal} instead of the WKB formula
that is used in Eq.~\ref{eq:FN1}. Consequently, the exact values are multiplied
by a  correction factor $\lambda$ to match the prediction of Eq.~\ref{eq:angular}.
For $E_0 = 3,6$ and $9$~MV/m the value of $\lambda$ is $1.273,1.15$ and $1.18$ respectively \cite{inspection}.
Note that Fig.~\ref{fig:angular} indicates that emission away from the apex (larger $\tth$) becomes significant
at higher applied fields.

The angular variation is thus reproduced exactly by Eq.~\ref{eq:angular}. Keeping in mind the
correction factor, the angular variation formula may be expressed as

\be
f_S(\tth) d\tth = \lambda 2\pi R_a^2~\frac{\sin\tth}{\cos^2\tth} \frac{A_\fn}{\phi}\frac{(E_a)^2}{t_F^2(\tth)} e^{-\frac{B_\fn \nuv_F(\tth) \phi^{3/2}}{(E_a \cos\tth)}} d\tth \label{eq:angular1}
\ee

\noi
where $\lambda$ accounts for the discrepancy between the exact current density and Eq.~\ref{eq:FN1}.
Note that when $f_S(\tth)$ is normalized for a probabilistic interpretation, the factor $\lambda$
is inconsequential.

The quantity $f_S(\tth)d\tth$ describes the distribution of electrons launched from the
surface of a parabolic emitter. Clearly, there is little emission along the axis and the
position where the peak occurs shifts away from the axis as the local apex field is increased.

Note that the distribution can be equivalently expressed in terms of $\rho$ and 
for sharp parabolic emitters ($R_a/h << 1$), $\rho \simeq R_a \tan\tth$. Using this relation
between $\rho$ and $\tan\tth$, it is easy to show that that the angle $\theta_L$ that
the normal (at any point $\rho$ on the parabolic surface $z = h - \rho^2/2R_a$)
makes with the emitter axis, is such that $\tan\theta_L = \rho/R_a \simeq \tan\tth$.
Thus, $\theta_L \simeq \tth$ so that
$f_S(\tth)$ also describes the distribution of launch angles ($\theta_L$) of electrons from the
surface of a parabolic emitter.

\section{Current from a single emitter}

The current emitted by a single sharp emitter tip can be calculated by integrating Eq.~\ref{eq:angular}
over $\tth$. Writing ${\cal A} =  2\pi R_a^2$ and collecting the constants (non-$\tth$ dependent terms)
together as

\be
\cc = \lambda {\cal A} \frac{A_\fn}{\phi} E_a^2
\ee

\noi
the current from the emitter can be expressed as

\be
I = \cc \int_0^{\tth_{max}} \frac{\sin\tth}{\cos^2\tth} \frac{1}{t_F^2(\tth)} e^{-\cb \nuv_F(\tth)/\cos\tth} d\tth  \label{eq:totI}
\ee

\noi
where $\cb = \frac{B_\fn \phi^{3/2}}{E_a}$. Note that  for sharp emitters, $\tth \simeq \pi/4$
corresponds to $\rho = R_a$ while $\tth = \pi/3$  corresponds to $\rho \simeq 2R_a$.
It is seen that the quadratic approximation is generally valid upto $\tth = \pi/3$.
Also, since the current beyond $\tth = \pi/3$ is extremely small,
the upper limit of integration can for all practical purposes be taken as $\tth_{max} = \pi/3$.
Thus,

\bea
I & = & \cc \int_0^{\pi/3} \frac{\sin\tth}{\cos^2\tth} \frac{1}{t_F^2(\tth)} e^{-\cb \nuv_F(\tth)/\cos\tth} d\tth  \\
& = & \cc \int_1^2 \frac{1}{t_F^2(u)} e^{-\cb \nuv_F(u) u} du
\eea

\noi
where we have used the substitution $u = 1/\cos\tth$. Since most of the emission occurs near the apex, it
is profitable to use the substitution $u = 1 + x$. Thus,

\be
I = \cc \int_0^1 \frac{1}{t_F^2(x)} e^{-\cb \nuv_F(x) (1 + x)} dx  \label{eq:totI1} 
\ee

\noi
where

\bea
\nuv_F(x) & = & 1 - \frac{f_0}{1 + x} + \frac{1}{6} \frac{f_0}{1 + x} \ln\big(\frac{f_0}{1 + x}\big) \\
t_F(x) & = & 1 + \frac{1}{9}\frac{f_0}{1 + x} - \frac{1}{18}\frac{f_0}{1+x}\ln\big(\frac{f_0}{1+x}\big) \\
f_0 & = &  c_S^2 \frac{E_a}{\phi^2}.
\eea

\noi
An expansion of $\nuv_F(x) (1 + x)$ and $1/t_F^2(x)$ in powers of $x$ is helpful in carrying out
the integration in Eq.~\ref{eq:totI1} since the dominant contribution is close to $x = 0$.
Retaining the first two terms yields,

\bea
\nuv_F(x) (1 + x) & = & D_0 + D_1 x + \mathcal{O}(x^2) \\
\frac{1}{t_F^2(x)} & = & F_0 + F_1 x +  \mathcal{O}(x^2)
\eea

\noi
where

\bea
D_0 & = & \nu_0 = 1 - f_0 + \frac{1}{6} f_0 \ln(f_0) \\
D_1 & = & 1 - \frac{1}{6} f_0 \\
F_0 & = & \frac{1}{t_0^2} = \frac{1}{(1 + \frac{f_0}{9} - \frac{f_0}{18} \ln f_0)^2} \\
F_1 & = & \frac{1}{9} \frac{ f_0 - f_0 \ln f_0}{(1 + \frac{f_0}{9} - \frac{f_0}{18} \ln f_0)^3}. 
\eea

\noi
Thus,

\bea
I & = &  \cc \int_0^1 (F_0 + F_1 x) e^{-\cb (D_0 + D_1 x)} dx  \nonumber\\
& \simeq & \cc e^{-\cb D_0} \Big[\frac{F_0}{\cb D_1} + \frac{F_1}{(\cb D_1)^2} - \frac{e^{-\cb D_1}}{\cb D_1}\big(F_0 + F_1 + \frac{F_1}{\cb D_1}\big)\Big] \nonumber \\
& = & 2 \pi R_a^2 J_{apex} \cg \label{eq:totI}
\eea

\noi
is the total current emitted by  a single emitter which can be expressed in terms of the apex current density $J_{apex}$,
the area ${\cal A} = 2\pi R_a^2$
of a hemisphere of radius $R_a$ and the area factor $\cg$ where

\be
\cg = \frac{1}{\cb D_1} + \frac{F_1}{F_0} \frac{1}{(\cb D_1)^2} - \frac{e^{-\cb D_1}}{\cb D_1}\big(1 + \frac{F_1}{F_0} + \frac{F_1}{F_0 \cb D_1}\big). \label{eq:gfact}
\ee

\begin{figure}[htb]
%\vskip -2.1 cm
%\hskip -1.0cm
%\centering
\hspace*{-1.0cm}\includegraphics[width=0.5\textwidth]{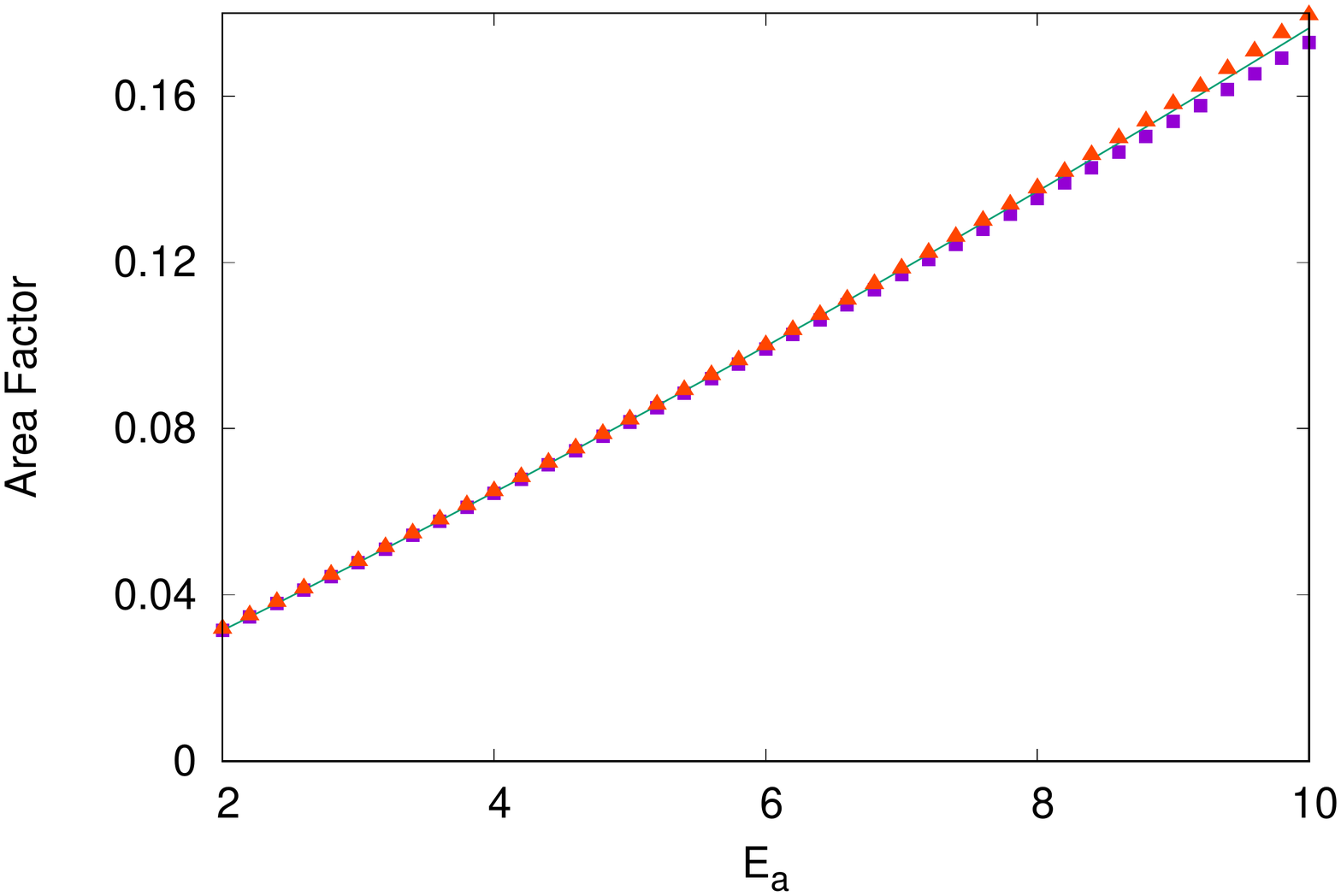}
\vskip -0.6 cm
\caption{The variation of the area factor $\cg$ with the apex field $E_a$ using  the exact transmission coefficient (triangle), using Eq.~\ref{eq:FN1} for the current density (squares) and Eq.~\ref{eq:gfact} (continuous line).
}
\label{fig:gfact}
\end{figure}

Eqns.~\ref{eq:totI} and \ref{eq:gfact} approximate the total current from an emitter and the effective emission area quite accurately as predicted by Eq.~\ref{eq:FN1}.
Fig.~\ref{fig:gfact} shows a plot of the area factor $\cg$ for local apex fields in the
range 2-10 V/nm using Eq.~\ref{eq:gfact} alongside the prediction using Eq.~\ref{eq:FN1} and the exact result using numerically determined transmission coefficient.
The agreement is excellent except at higher apex fields where the analytical result (based on Eq.~\ref{eq:FN1}) is closer to the exact result. Note that the factor $\lambda$ is
immaterial as far as the area factor is concerned.

For materials with workfunction in the range 4-5 eV and apex fields in the range 1-10 V/nm, the term $e^{-\cb D_1}/(\cb D_1)$
is negligible. A simple and reasonably accurate formula for the area factor is thus

\bea
\cg & \simeq &  \frac{1}{\cb D_1} + \frac{F_1}{F_0} \frac{1}{(\cb D_1)^2} \\
& = & \frac{E_a}{B_{\fn} \phi^{3/2}}  \frac{1}{(1 - f_0/6)} \Big[ 1 + \frac{F_1}{F_0} \frac{1}{(\cb D_1)} \Big] \nonumber
\eea

\noi
where the second term in the square bracket is a small correction even at high apex fields and may be neglected.
For small $E_a$ therefore, $\cg \sim E_a$ \cite{henderson}.

\section{The total energy distribution}

The joint distribution of emitted electrons $f_J$ with respect to the quantities $\tth$, the normal energy  $\ce_N$ and the
total energy $\ce_T$ can be expressed as

\be
\begin{split}
  f_J(\tth,\ce_N,\ce_T)  d\tth d\ce_N d\ce_T = \Big[\ca   \frac{\sin\tth}{\cos^4\tth}
      & n(\ce_T) D(\ce_N,\tth) \Big] \\ & d\tth d\ce_N d\ce_T.  \label{eq:joint}
  \end{split}
\ee

\noi
where the supply function density

\be
n(\ce_T) = \frac{2em}{(2\pi)^2\hbar^3} f_{FD}(\ce_T,T)
\ee

\noi
and  $f_{FD}(\ce_T,T)$ is the Fermi-Dirac distribution at a temperature $T$.
By integrating over any two of these quantities, the distribution over a third quantity can be determined.
It can also be used to arrive at joint distribution of any two of ($\tth,\ce_N,\ce_T$) by integrating over the third. In the
rest of this paper, we shall be interested in the distribution of emitted electrons with respect to the
total and normal energy . For a given normal energy, the total energy ranges from $\ce_N$ to infinity (or $\ce_F$
at zero temperature) while the normal energy component which determines
the transmission coefficient, varies from 0 to the total energy $\ce_T$. The total energy distribution
of the current can thus be expressed as

\be
f_T(\ce_T) d\ce_T = \Big[\ca \int \frac{\sin\tth}{\cos^4\tth} d\tth \int_0^{\ce_T} n(\ce_T) D(\ce_N,\tth) d\ce_N\Big] d\ce_T
\ee

\noi
where $\ce$ and $\ce_N$ are the total and normal energy respectively, $D(\ce_N)$ is the transmission coefficient at a normal energy $\ce_N$:

\be
D(\ce_N,\tth) = e^{-\nu_F(\cos\tth) \cb /\cos\tth} e^{-(\ce_F - \ce_N) t_F(\cos\tth)/d_F\cos\tth}
\ee

\noi
with  $1/d_F = g_e \phi^{1/2}/E_a$ and $g_e = \sqrt{8m/\hbar^2} \simeq 10.24634~{\rm eV}^{-1/2} {\rm nm}^{-1} $.

The integration over the normal energy $\ce_N$ can be carried over $(0,\ce_T)$ to yield the joint
distribution  $f(\tth,\ce_T)$

\be
f(\tth,\ce_T) = \ck  e^{-\cb \nu_F /\cos\tth} \frac{\sin\tth}{\cos^3\tth} \frac{e^{-(\ce_F - \ce_T) t_F/d_F\cos\tth}}{t_F}
\ee

\noi
and on integrating $f(\tth,\ce_T)$ over $\tth$, the total energy distribution of emitted electrons can be
obtained as

\be
f_T(\ce_T) =  \ck \int_0^{\pi/3} d\tth  e^{-\cb \nu_F /\cos\tth} \frac{\sin\tth}{\cos^3\tth} \frac{e^{-(\ce_F - \ce_T) t_F/d_F\cos\tth}}{t_F}  \label{eq:totP1}
\ee

\noi
where $\ck = \ck_0 f_{FD}(\ce_T,T)  d_F$,  $\ck_0  =  \frac{2em}{(2\pi)^2\hbar^3} 2\pi R_a^2$
and $\nu_F$ and $t_F$ are functions of $\cos\tth$.

\begin{figure}[htb]
%\vskip -2.1 cm
%\hskip -1.0cm
%\centering
\hspace*{-1.0cm}\includegraphics[width=0.55\textwidth]{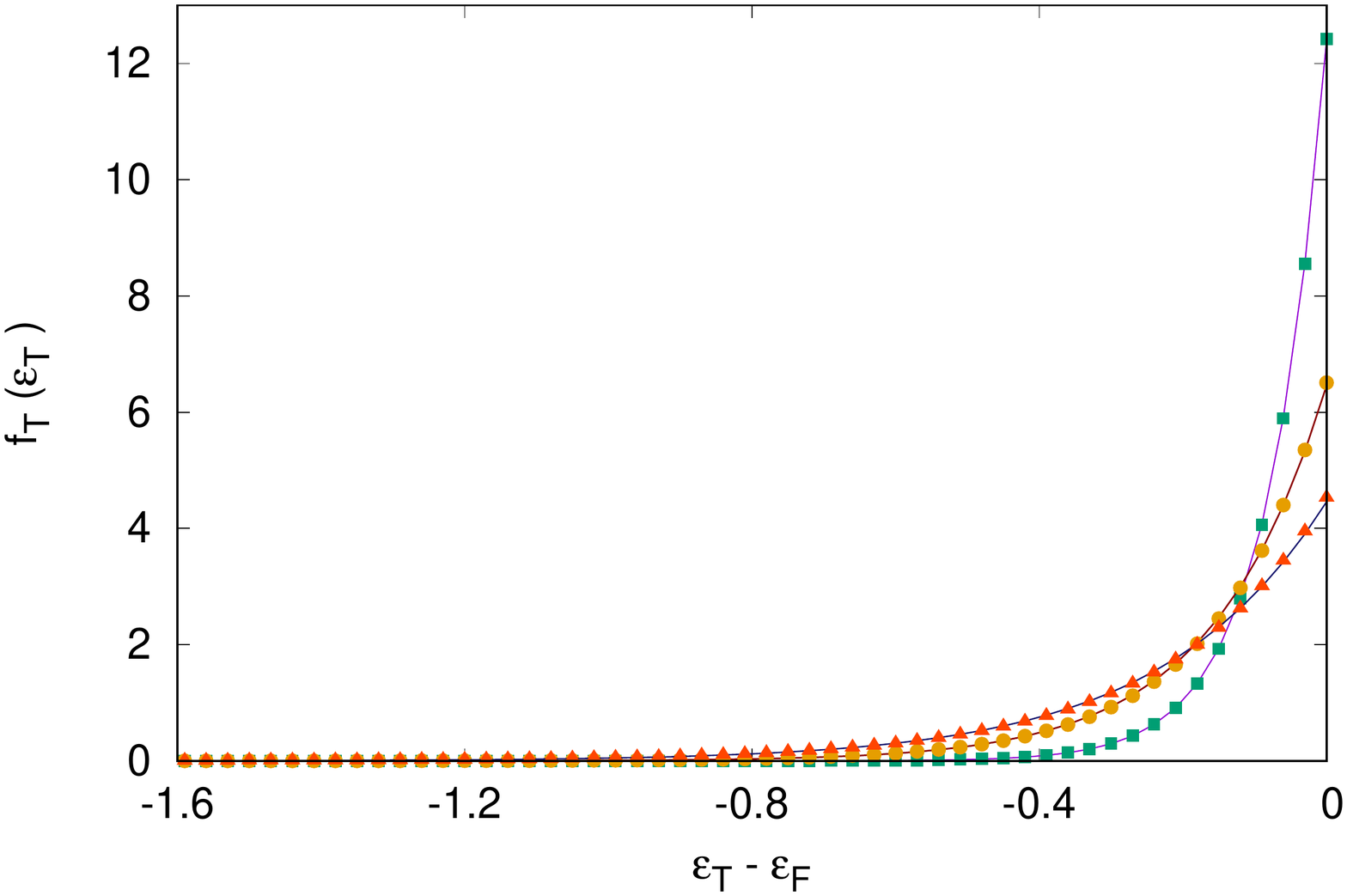}
\vskip -0.6 cm
\caption{The normalized total-energy density $f_T(\ce_T)$ of Eq.~\ref{eq:totE} for the system described in Fig.~\ref{fig:angular} at zero temperature for (a) $E_a = 1.85$~V/nm (square) (b) $E_a = 3.70$~V/nm (circle) and (c) $E_a = 5.55$~V/nm (triangle). The continuous lines are the corresponding exact results using numerically determined transmission coefficients.
}
\label{fig:totalE}
\end{figure}

Using the substitutions $\cos\tth = 1/u$ and 
and $u = 1 + x$, Eq.~\ref{eq:totP1} simplifies as

\be
f_T(\ce_T) =  \ck \int_0^1 dx  e^{-\cb \nu_F(x) (1+x)} \frac{1+x}{t_F(x)} e^{-(\ce_F-\ce_T) \frac{t_F(x)}{d_F}(1+x)}  \label{eq:totP2}
\ee

\noi
The integral can be carried out approximately using the expansions:

\bea
\nu_F(x) (1 + x) & = & D_0 + D_1 x \\
t_F(x) (1 + x) & = & G_0 + G_1 x \\
\frac{1+x}{t_F(x)} & = & H_0 + H_1 x
\eea

\noi
where $D_0$ and $D_1$ have been defined before and

\bea
G_0 & = & t_0 = 1 + \frac{f_0}{9} - \frac{f_0}{18} \ln f_0  \\
G_1 & = & 1 + \frac{f_0}{18} \\
H_0 & = & \frac{1}{1 + \frac{f_0}{9} - \frac{f_0}{18} \ln f_0 } = \frac{1}{G_0} \\
H_1 & = & \frac{1 + \frac{f_0}{6} - \frac{1}{9} f_0 \ln f_0 }{\big(1 + \frac{f_0}{9} - \frac{1}{18} f_0 \ln f_0\big)^2}.
\eea
    
The integrations can now be carried out so that the total energy probability density reduces to

\be
f_T(\ce_T) = \ck e^{-\delta_0} \Big[\big(H_0 + \frac{H_1}{\delta_1}\big)\frac{1 - e^{-\delta_1}}{\delta_1} - \frac{H_1}{\delta_1} e^{-\delta_1}\Big] \label{eq:totE}
\ee

\noi
where

\bea
\delta_0 & = & \cb D_0 + (\ce_F - \ce_T) G_0/d_F \\
\delta_1 & = & \cb D_1 + (\ce_F - \ce_T) G_1/d_F.
\eea

\noi
The total energy distribution can thus be expressed in terms of the apex electric field $E_a$
and radius of curvature $R_a$.

The total energy distribution of Eq.~\ref{eq:totE} is compared with the exact result obtained
using numerically determined transmission coefficients in Fig.~\ref{fig:totalE} at zero
temperature. The density $f_T(\ce_T)$ is normalized in both
cases so that the correction factor $\lambda$ is
immaterial. The agreement is excellent for the range of field strengths considered.
Also, the longer tail at higher applied fields indicates larger contributions to the
current from lower energy electrons.
Thus, Eq.~\ref{eq:totE} serves as a good
approximation for the total energy distribution and can be used to determine the
peak position (also at non-zero temperature) with respect to the local field at the apex and the work function
\cite{liang}.

As in case of the area factor, the distribution $f_T(\ce_T)$ can be simplified considerably, for typical
fields and work function of interest in field emission, without compromising much with its accuracy.
A good first approximation that is reasonably accurate is

\bea
f_T(\ce_T) & \simeq & \ck H_0 \frac{e^{-\delta_0}}{\delta_1} \label{eq:fT_approx} \\
& = &  \frac{{\cal A}}{\delta_1} \frac{2em}{(2\pi)^2\hbar^3} \frac{d_F}{t_0} e^{-\frac{B_{FN}\phi^{3/2}}{E_a} \nu_0} e^{-(\ce_F - \ce_T)t_0/d_F} \nonumber
\eea

\noi
which can be compared with the total energy distribution of the current density at the apex

\be
f_T^{apex} (\ce_T) = \frac{2em}{(2\pi)^2\hbar^3} \frac{d_F}{t_0} e^{-\frac{B_{FN}\phi^{3/2}}{E_a} \nu_0} e^{-(\ce_F - \ce_T)t_0/d_F} . \label{eq:fT_old}
\ee

\noi
Since the two densities are unnormalized,  the difference in this simplified picture lies essentially
in the

\be
\frac{1}{\delta_1} = \frac{1}{\frac{B_{FN} \phi^{3/2}}{E_a} (1 - f_0) + \frac{(\ce_F - \ce_T)}{d_F}(1 + f_0/18)}
\ee
  
\noi
term. As a measure of the difference, the mean energy of the emitted electrons can be
calculated. For the total energy distribution of the current density at the apex, we
get the well known result \cite{swanson}

\be
\langle \ce_T^{apex} \rangle = \ce_F - \frac{d_F}{t_0}
\ee

\noi
while for the total energy distribution of the net current from a single emitter (Eq.~\ref{eq:fT_approx})

\be
\langle \ce_T \rangle  =  \ce_F - \frac{d_F}{t_0} \frac{\Gamma(-1,\alpha)}{\Gamma(0,\alpha)} 
\ee

\noi
where

\be
\alpha = \frac{B_{\fn} \phi^{3/2}}{E_a}\frac{t_0 (1 - f_0/6)}{1 + f_0/18}. \label{eq:alpha}
\ee

\noi
A comparison shows that the difference in mean electron energies is about $10\%$ for fields
exceeding $5$~V/nm with $\langle E_T \rangle$ being higher than $\langle E_T^{apex} \rangle$.

\section{The normal energy distribution}

The normal energy distribution can similarly be obtained by integrating $f_J$ over $\tth$ and $\ce_T$. It is a
quantity of interest in its own right and can be used in determining conditional distributions \cite{egorov}.
It can be expressed as

\be
f_N(\ce_N)  d\ce_N  =  \Big[ \ca \int d\tth \frac{\sin\tth}{\cos^4\tth} D(\ce_N,\tth)  \int_{\ce_N}^{\infty} n(\ce_T)  d\ce_T \Big] d\ce_N.
\ee

\noi
The integration over the total energy can be performed easily and with the approximation
$k_BT \ln\big(1 + e^{\frac{\ce - \ce_F}{k_BT}}\big) \simeq \ce_F - \ce_N$, the expression for the
normal energy probability density takes the form

\be
f_N(\ce_N) =   \ck_0  (\ce_F - \ce_N) \int_0^{\pi/3} \frac{\sin\tth}{\cos^4\tth} D(\ce_N,\tth)  d\tth \label{eq:normE}
\ee

\noi
where the transmission coefficient $D(\ce_N)$ depends on $\tth$. The joint distribution
$f(\ce_N,\tth)d\ce_N d\tth$ is thus

\be
f(\ce_N,\tth) =   \ck_0  (\ce_F - \ce_N) \frac{\sin\tth}{\cos^4\tth} D(\ce_N,\tth)
\ee

\noi
where $f(\ce_N,\tth) d\ce_N d\tth$ measures the current with normal energy between $\ce_N$ and
$\ce_N + d\ce_N$ and (generalized) angle between $\tth$ and $\tth + d\tth$.

\begin{figure}[htb]
%\vskip -2.1 cm
%\hskip -1.0cm
%\centering
\hspace*{-1.0cm}\includegraphics[width=0.55\textwidth]{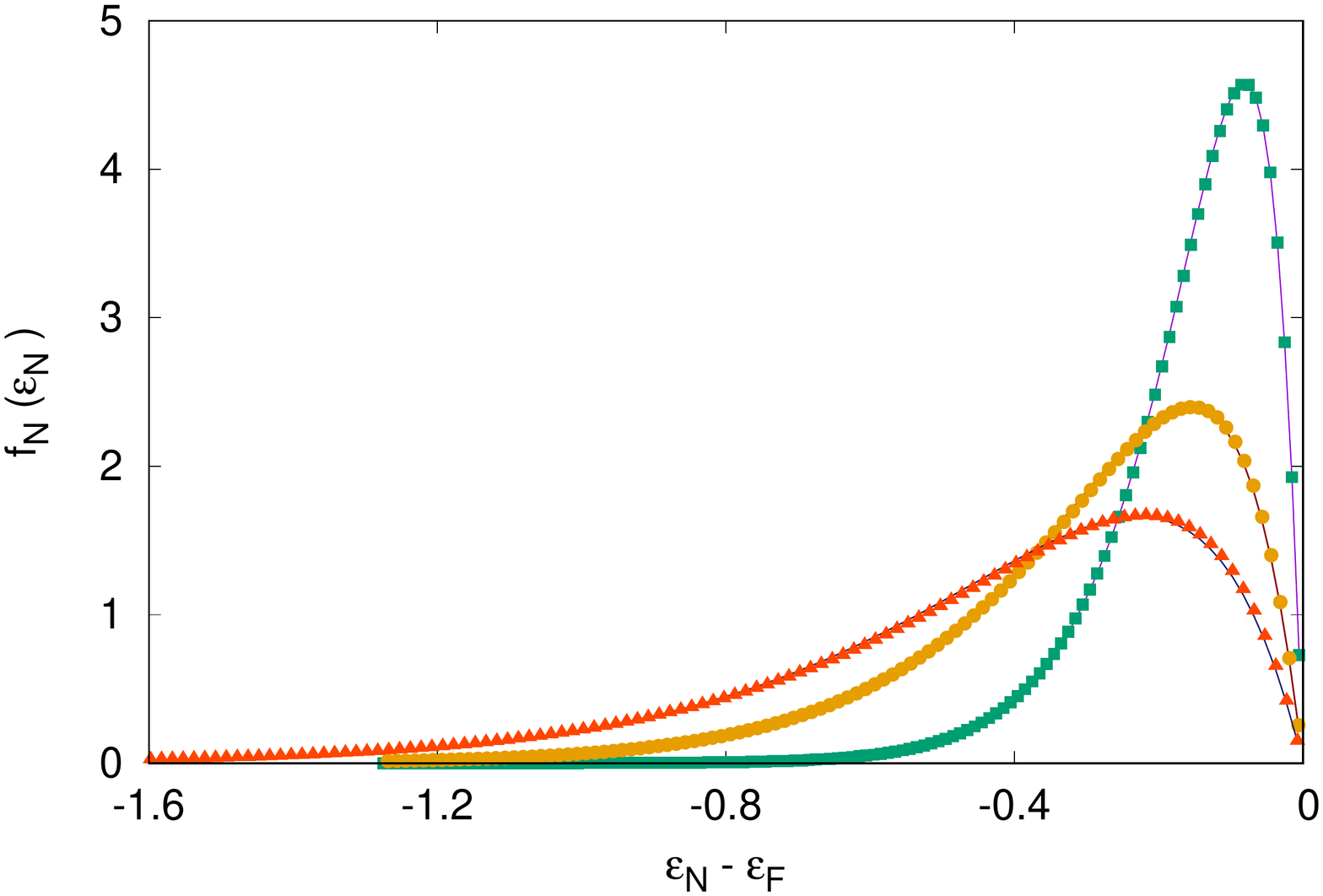}
\vskip -0.6 cm
\caption{The normalized normal-energy density $f_N(\ce_N)$ of Eq.~\ref{eq:f_N} for the system described in Fig.~\ref{fig:angular} at zero temperature for (a) $E_a = 1.85$~V/nm (square) (b) $E_a = 3.70$~V/nm (circle) and (c) $E_a = 5.55$~V/nm (triangle). The continuous lines are the corresponding exact results using numerically determined transmission coefficients.
}
\label{fig:normalE}
\end{figure}

Using the transformations
$u = 1/\cos\tth$ and $u = 1 + x$ and the expansions for $\nu_F(x) (1+x)$ and $t_F(x) (1+x)$,
Eq.~\ref{eq:normE} can be expressed as

\be
\begin{split}
  f_N(\ce_N) = \ck_0 (\ce_F - \ce_N) & \int_0^1~e^{-\cb (D_0 + D_1 x)} (1+x)^2 \times \\
    & e^{-\frac{(\ce_F - \ce_N)}{d_F} (G_0 + G_1 x)} dx.
\end{split}
\ee

\noi
The integrations can be performed to yield 

\be
\begin{split}
  f_N(\ce_N) =  \ck_0 e^{-\delta_2} & (\ce_F - \ce_N) \Big[ \big(1-e^{-\delta_3}\big)\big(\frac{1}{\delta_3} + \frac{2}{\delta_3^2} +  \\ &
    \frac{2}{\delta_3^3}\big) - 
     \frac{3}{\delta_3}e^{-\delta_3}  -  \frac{2}{\delta_3^2}e^{-\delta3} \Big]
  \end{split} \label{eq:f_N}
\ee

\noi
where $\ck_0$ and

\bea
\delta_2 & = & \cb D_0 + (\ce_F - \ce_N) G_0/d_F \\
\delta_3 & = & \cb D_1 + (\ce_F - \ce_N) G_1/d_F.
\eea

\noi
Eq.~\ref{eq:f_N} is found to be a good approximation
of the normal-energy distribution of emission current as shown in Fig.~\ref{fig:normalE}.
As seen before, with an increase in applied field, the spread in the normal energy distribution increases
and the peak shifts away from the Fermi energy. Eq.~\ref{eq:f_N}
can be used to determine other quantities of interest such as the
peak position at non-zero temperature with respect to the local field at the apex and the work function.

A simple but accurate approximation can again be made as in case of the total energy distribution.
To a reasonably good approximation,

\bea
f_N(\ce_N) & \simeq & \ck_0 e^{-\delta_2} (\ce_F - \ce_N) \frac{1}{\delta_3} \\
& = & \frac{\ck_0}{\delta_3}  (\ce_F - \ce_N) e^{-\frac{B_{FN}\phi^{3/2}}{E_a} \nu_0} e^{-(\ce_F - \ce_N)\frac{t_0}{d_F}} \nonumber
\eea

\noi
which can be compared with the normal energy distribution of the current density at the apex

\be
f_N^{apex}(\ce_N) = \frac{\ck_0}{\cal A} (\ce_F - \ce_N) e^{-\frac{B_{FN}\phi^{3/2}}{E_a} \nu_0} e^{-(\ce_F - \ce_N)\frac{t_0}{d_F}} .
\ee

\noi
Since the densities are unnormalized, the difference essentially lies in the term

\be
\frac{1}{\delta_3} = \frac{1}{\frac{B_{FN} \phi^{3/2}}{E_a} (1 - f_0) + \frac{(\ce_F - \ce_N)}{d_F}(1 + f_0/18)}.
\ee

\noi
As in case of the total energy, the mean normal energies can be calculated in order to compare the
two distributions. Thus, we have

\be
\langle \ce_N^{apex} \rangle = \ce_F - 2 \frac{d_F}{t_0}
\ee

\noi
while

\be
\langle \ce_N \rangle = \ce_F - 2 \frac{d_F}{t_0} \frac{\Gamma(-2,\alpha)}{\Gamma(-1,\alpha)}
\ee

\noi
where $\alpha$ is defined in Eq.~\ref{eq:alpha}. The difference for $E_a > 5$~V/nm is again about
$10\%$ with $\langle \ce_N \rangle$ higher than $\langle \ce_N^{apex} \rangle$.

\section{Summary and Discussions}

We have derived analytical expressions for total field-emission current, its distribution
on the emitter surface in terms of generalized angle, $\tth$ (which coincides with the
distribution of launch angles) as well the distributions with
respect to the total and normal energies. In the process, we also provide joint distributions
of ($\tth,\ce_T$) and ($\tth,\ce_N$) and an expression for the effective emission area.
All of these are based on a recent result on the variation of field enhancement factor
near the apex of a smooth emitter. Despite the approximations used, the expressions
are in good agreement with exact results where the transmission coefficients
are evaluated numerically.

The results presented here are strictly valid when the emitter has a uniform
work function over the emission surface and when it is placed on a flat surface with its axis
parallel to an asymptotic uniform electrostatic field. The result may
continue be of some significance when these conditions are somewhat relaxed.

An accurate use of these expressions for making useful predictions for experiments is however limited by 
our sketchy knowledge about the apex field enhancement factor, $\gamma_a$ or alternately
the electric field at the apex, $E_a$. A recent advancement in this direction \cite{db_fef}
is likely to bring theoretical predictions closer to experimental results at least for
single emitters.

\section{Acknowledgments}

The author acknowledges several useful discussions with Rajasree, Gaurav Singh and Raghwendra Kumar
and thanks them for a critical reading of the manuscript.

\vskip 0.05 in
$\;$\\
\section{References} 

%\begin{references}


\begin{thebibliography}{99}
\bibitem{uncertainties} Even when these are known (e.g. for a hemi-ellipsoidal emitter), there may be other
experimental uncertainties such as surface finish or adsorbed gases that hinder reasonable theoretical predictions. However
we shall steer clear of such complications and merely concentrate on an ``ideal'' theoretical prediction
for generic emitters.
\bibitem{emperical} Various emperical formulae for the apex field enhancement factor $\gamma_a$
  exist, all of which are expressed in terms of the height $h$ and the apex radius of curvature $R_a$.
\bibitem{sharp_tip}  Recent results have shown that curvature corrections
to the image and external potential are important when $R_a < 20$~nm and especially so
when $R_a < 5$~nm.
\bibitem{inspection} The values of $\lambda$ follow the trend arrived at by Mayer \cite{Mayer}.
A recipe for determining the discrepancy factor between the transfer-matrix (TM)
and Murphy-Good (MG) formalism at arbitrary $\phi$, $\ce_F$ and the local external field $E_a$
is given in Table 1 of Mayer \cite{Mayer}.
\end{thebibliography}
\end{document}